\documentstyle[prb,aps]{revtex} 
\newcommand{\bu}{\mbox{\boldmath $\hat{B}$}}
\newcommand{\ru}{\mbox{\boldmath $\hat{R}$}}
\newcommand{\r}{\mbox{\boldmath $R$}}
\newcommand{\M}{\mbox{\boldmath $M$}}
\newcommand{\be}{\begin{equation}}
\newcommand{\ee}{\end{equation}}
\newcommand{\sch}{\mbox{$\mathcal{H}$}}
\begin{document}
\input epsf
\def\eg{{\it e.g.}}
\def\ie{{\it i.e.}}
\def\nn{n.n.}
\def\nnn{ n.n.n.}
\def\etal{{\it et\,al.}}
%%%%%%%%%%%%%%%%%%%%%%%%
\draft
\twocolumn[\hsize\textwidth\columnwidth\hsize\csname
@twocolumnfalse\endcsname

% *** Title Page **************************************************************
\title{Bond-Ordering Model for the Glass Transition}

\author{Saeid Davatolhagh\cite{saeid} and  Bruce R.\ Patton\cite{bruce} }

\address{Department of Physics, The Ohio State University, Columbus, Ohio 43210}

\date{ \today}

\maketitle

\begin{abstract}
We consider the idea of bond ordering as a model for the
glass transition; a generic covalently bonded liquid may substantially
reduce its energy through bond ordering, without undergoing 
significant structural order. This concept is developed for a model system using
quantities such as a bond order-parameter and  susceptibility which  
provide a new identification  for the calorimetric 
glass transition temperature. Monte Carlo simulation results exhibit bond ordering 
at intermediate temperatures uncorrelated with any long-range structural ordering. Also discussed are various 
other implications of bond-ordering model for the glass transition.
\end{abstract}
\vskip2pc]
\section{Introduction}
The problem of glass formation is a classic one in physics and materials science.
Viewpoints break down into kinetic explanations of an arrested liquid, and phase 
transitions with some kind of order parameter. Discussions have approached the glass transition 
in terms of the fragmentation of small structural units,\cite{0.1}  
the agglomeration of clusters through chemical bonds,\cite{0.2} or 
correlations between different metastable equilibrium states representing 
distinct configurations or rearrangements of the system.\cite{0.3}
In addition, attempts have been made to incorporate frustration in local 
bond ordering in a glass through the introduction of a local order parameter 
describing locally preferred arrangements of liquid molecules, which, in general,
are not consistent with the crystallographic symmetry favored by
density ordering.\cite{0.4}  In this model the frustration arises from competition between 
density ordering and local bond ordering, explaining why some molecules crystallize easily without vitrification, 
while others easily form glasses without crystallization.  Another approach, using computer 
simulations, introduces a displacement--displacement correlation function as a measure of 
the local configurational order which grows as the temperature is lowered toward the 
glass transition.\cite{0.5}  Finally, the role 
of fluctuations has been reviewed in general fashion with both theoretical and experimental 
evidence for heterogeneity at the glass transition.\cite{0.6} 
We would like to introduce a model of  positional glass which incorporates 
many of these ideas in simple form.

First we recall some of the basic structural features of typical network-glass-formers like silica. 
Vitreous  silica is made up of SiO$_4$ tetrahedra just like the crystalline phase, quartz. Thus,
molecular structure for the two different phases of silica, \ie\ crystalline and amorphous,
are very much alike, with the major exception of \nnn\ distance Si$_1$-Si$_2$
which may be  specified in terms of the Si-O-Si bond angle.
The distribution of Si-O-Si bond angles $\beta$ as  determined by Mozzi and Warren
for vitreous silica is shown in Fig.\,1.\cite{1} From the bond angle distribution V($\beta$),
it is  clear that the Si$_1$-Si$_2$ distances vary significantly,
which in fact is the primary source of  topological disorder. Yet the bond angle $\beta$
is confined to  within roughly $30^o$ of its most probable value  {\it viz} $144^o$.
Clearly, the  variation in $\beta$ is  large enough to suppress
the long-range order (LRO) characterizing  the crystalline phase, and yet small enough to maintain the
medium-range  order (MRO)  which   extends to about $10\AA$.
Hence, we note that ordering in  \nn\ and \nnn\ distances  or bond lengths is well 
maintained in spite of the fact that vitreous silica is devoid of 
any substantial structural order. Similar structural properties are observed 
in various other network glasses such as B$_2$O$_3$ and GeO$_2$.
This observation is striking enough to indicate a strong role for the ordering of bonds in 
a proper microscopic model for the glass transition. The bond ordering can be viewed 
as being brought about by the
local reorientations of molecular clusters as suggested in the theory of 
Adam and Gibbs.\cite{2}
The idea is that  supercooled liquids do not necessarily need to undergo structural ordering 
in order to achieve local equilibrium. In fact, local rearrangements of   molecular clusters can lead
to substantial lowering in the internal energy of an entire sample through reducing bond energies at
the local level, which is the primary reason for the ordering of 
\nn\ and \nnn\ distances, or alternatively the bond-angle degrees of freedom, in amorphous materials.  

A  model for the glass transition incorporating small structural units or fragments 
was proposed by M.~Suzuki \etal\cite{0.1}. In this fragmentation model noncrystalline 
solids are assumed to be assemblies of pseudo-molecules---a pseudo-molecule being a cluster of atoms
having a disordered lattice in which there are no definite defects such as under- or over-coordinated
atoms. As temperature increases, bond breaking intensifies at the boundaries of such clusters
where the bonds tend to be weak. The bond breaking mechanism arising from the thermal excitation of 
electrons from bonding to antibonding energy states, causes the noncrystalline
solid to fragmentize, with the average fragment size 
decreasing as the temperature increases. Consequently, material begins to show viscous
flow when average fragment size reaches a critical value. The model is shown to have 
some success in describing the temperature dependence of viscosity and  the variation in  
transition temperature with heating rate for \mbox{a-Si}. The origin of pseudo-molecules in the 
cooling process, however, is not addressed in the fragmentation model.

In the following we consider a microscopic model for the glass transition which is general enough to be applicable to 
various types of glass-forming systems with widely differing bonding schemes and chemical compositions. 
We would like to treat glass transition as a phenomenon that finds similar description whether a liquid is 
cooled or heated through the transition.
The focus of  bond-ordering model is the bonds linking neighboring atoms rather than the 
atoms themselves. In other words, a bond is treated as a distinct object in its own right, possessing
internal  degrees of freedom or electronic states. The internal state of a bond is 
governed by the separation of the participating atoms. 
The term bond-ordering refers to the  process of relaxation of bonds into
their low-lying internal energy states, facilitated by the cooperative rearrangements within molecular
groups. Bond ordering, therefore, may be viewed as some form of  ordering in energy space. 
The important point that 
we would like to bring home, however, is that such an ordering can be achieved without need for any 
significant structural order. To this effect, we have provided results from Monte Carlo (MC) 
simulations of the model Hamiltonian which couples the coordinates of ions to the electronic states of 
electrons, as for a typical covalently bonded network glass. The simulations make clear the possibility 
of local ordering of bonds,  uncorrelated with any kind of long-range structural ordering.

Sec.\,II\@ of this article contains the theoretical background regarding the covalently-bonded systems. 
In Sec.\,III we describe the model Hamiltonian.
Some new definitions in terms of mathematical expressions for various 
structural and bond-related quantities of interest, are introduce in Sec.\,IV\@.
The simulation procedure and results are discussed 
extensively in Sec.\,V\@. Sec.\,VI contains the  concluding remarks and a summary of the main ideas 
introduced in this article.

\section{Theoretical Background}
Glass-forming liquids include covalently-bonded network glasses, coordination-constrained metallic
glasses and systems made of more complex organic molecules or polymer. Here we take the view that 
the interaction between constituent atoms or molecular units is given by an effective potential 
which characterizes their positional and orientational interaction. In particular for covalent materials 
the interactions may be reasonably described in terms of basis states of degenerate $s$ and $p$ orbitals on
each atom. Linear combinations of orbitals on neighboring atoms then lead to bonding and antibonding
states which describe the interactions between atoms. The occupation of the bonding state by one to two
electrons results in a lowering of the electronic energy relative to the atomic levels. This covalent bond 
provides the structural constraint on the relative position of the two atoms. 

In order to be more specific we consider a general Hamiltonian for a covalently bonded network glass.
We start first with a general formulation of the Hamiltonian for a typical glass-forming liquid. 
Although these systems are often treated in a classical formulation, in fact, the bonding between atoms in 
a network glass, as well as the coordination in metallic glasses arise from quantum mechanical considerations.
The many-body Hamiltonian may be written in the form:
\be
H=H_{ii}+H_{ie}+H_{ee}\;,
\ee
where $H_{ii}$ is the interaction between ions in the liquid, $H_{ie}$ is the ion-electron interaction,
and $H_{ee}$ is the interaction between electrons. The usual approach to solving this type of many-body 
problem is to exploit the separation of time scales inherent in the Born-Oppenheimer approximation: solve 
for the electronic states regarding the ionic coordinates as parameters, then vary the ionic coordinates 
to minimize the energy or extract the electron-phonon interaction, etc. For a covalent material like most
network glasses, the electronic states are reasonably approximated by forming bonding and antibonding states
from the atomic $s$ and $p$ orbitals. The available electrons are then apportioned to the lowest bonding states
to obtain the ground state of the system. The bonding and antibonding energies, depend on distance or separation
as shown schematically in Fig.\,1: the bonding state has a minimum at an ion-ion distance $r_0$, while the 
antibonding state is repulsive  at all distances. This may be thought of as a tight-binding approximation
to the actual electronic structure of a glass. A bond, is a bonding state occupied by two electrons and is 
strongest (or has lowest energy) when the distance between ions is near $r_0$. On the other hand, a bond missing
an electron due to a thermal fluctuation or a transition of the electron to the antibonding state, corresponds
to a broken bond. In this tight-binding representation of bonds, the Hamiltonian may be expressed
as in Eq.\,(1), a more detailed expression of which is derived in the next section.  

\section{Model Hamiltonian}
The potential energy of a pair interaction may be expanded in the displacement $x$ from the equilibrium:
\be
U(x)=U_0+\frac{x^2}{2!} U''_0+\frac{x^3}{3!} U'''_0+ \cdots\;,
\ee
where the coefficients are evaluated at equilibrium separation. 
Dropping the constant term which plays no role, and in the harmonic approximation, we get:
\be
U(x) =  \frac{x^2}{2} U''_0\; . 
\ee
For our lattice model we identify $x=| \r_i-\r_j |$, where $\r_i$ and $\r_j$
are the displacements from the equilibrium of the  atoms assigned to the $i$th
and $j$th \nn\  sites, respectively.
Hence, with the idealizing constraint that all displacements are of  same magnitude $|\r |$, 
\ie\ $|\r_i|=|\r_j|=|\r |$, we have:
\be
U\left(|\r_i-\r_j|\right)= |\r |^2\;U''_0 \left( 1-\ru_i\cdot\ru_j \right). 
\ee
Letting $J= |\r |^2\;U''_0$, we obtain the following Hamiltonian for a system of 
atoms interacting through \nn\ coupling:
\be
\tilde{\sch} = -J\sum_{<i,j>} \left( \ru_i \cdot \ru_j \right)\;.  
\ee
This is clearly of the same mathematical form as the \mbox{$q$-state} clock 
model Hamiltonian, when the displacement degrees of freedom are taken to be discrete.

The interaction Hamiltonian we used for our MC simulations, is the following:
\begin{eqnarray}
\sch & = & -J \sum_{<i,j>} \left( \ru_i \cdot \ru_j + 1\right)\;n_{ij}\nonumber \\
     & = & -J \sum_{<i,j>} \left( \cos \theta_{ij} + 1 \right)\;n_{ij}\;,
\end{eqnarray}
where $\theta_{ij} = \frac{2\pi}{8} (s_i-s_j)$, with $s_i,s_j =1,2,\ldots,8$.
The $s_i$'s  are integer labels for various possible displacements 
of an atom from its  equilibrium lattice site $i$\,.
The quantity  $n_{ij}$ should be regarded as the bonding-electron  occupation number
for a (possible) bond, linking  \nn\ sites $i$ and $j$. A bond may or may not be broken 
depending on  whether the corresponding $n_{ij}$ takes on the values $0$, or~$1$, respectively.
The value taken by  $n_{ij}$  depends on the number of bonding-electrons made 
available to the system---this is an input parameter to  the  simulation code---and the  relative values 
of energies of the \nn\ interactions. Bonds with lower values of energy are more likely to have  bonding-electrons.
Interesting effects are observed with a bonding-electron (hole) concentration  of about $60\%$ ($40\%$) 
and that is what we report later in this article.
As we shall see shortly, the net effect of  holes is  to suppress the $XY$-like transition 
occurring  for clock models with $q>4$, and hence allowing us to observe the
behavior of a disordered system with lowering of  the temperature.

\section{Some New Definitions}
In Sec.\,I we defined bond ordering as that process involving  relaxation of bonds 
into their low-lying internal energy states. There are a few more physical parameters 
relevant to our discussion which are defined in this  section and are considered in 
context of the  glass transition phenomenon.

\subsection{Bond Order-Parameter}
We introduce  bond order parameter (or bond magnetization) as a measurable physical property
that indicates the extent of bond ordering prevailing in a physical system.
We begin with considering a two dimensional  system of $4$-fold coordinated atoms interacting through 
\nn\ coupling of strength $J$. The bond magnetization of such a system is  significant
when  bonds  are in their low-lying 
energy states (as for a bond-ordered low temperature phase), and negligible
when bonds are  distributed among all  possible energy states with uniform probability 
which is indeed the case when the thermal
energy is far in excess of the coupling strength $J$. To construct an
expression for the  bond magnetization of such a system,
every \nn\ pair of atoms is characterized by a vector $\bu_{ij}$ 
the  purpose of which is to characterize the interaction energy.
$\bu_{ij}$ is specified via the angle  it makes with
an arbitrary fixed axis in the plain  of the system, $\phi_{ij}$. The angle $\phi_{ij}$ is written 
in terms of bond energy as:
\be
\phi_{ij}=\pi \frac{\epsilon_{ij}}{J}\;, \hspace{0.8cm}   -J\leq \epsilon_{ij} \leq J\;, 
\ee
where $J$ is \nn\ coupling strength, $\epsilon_{ij}$ is the bond energy and $-\pi \leq \phi_{ij} \leq \pi$\,.
An expression that fulfills all the requirements of an extensive bond magnetization
is the following:
\begin{eqnarray}
M_b & = & \left\langle \left(\sum_{<i,j>}^{2N} \bu_{ij}\right)^2\right\rangle^{1/2} \nonumber \\
    & = &\left( \sum_{<i,j>}^{2N} \sum_{<m,n>}^{2N} \left\langle \bu_{ij} \cdot 
\bu_{mn} \right\rangle \right)^{1/2}\;,
\end{eqnarray}
where angular braces stand for thermal average and $N$ is the total number of atoms in the system.
In terms of  bonds' energies, the expression for intensive bond magnetization of the system;
$m_b=M_b/2N$, may be written as:
\be
m_b=\frac{1}{2N}\left( \sum_{<i,j>}^{2N}\!\sum_{<m,n>}^{2N} \left\langle \cos\left[\frac{\pi}{J}
(\epsilon_{ij}-\epsilon_{mn})\right] \right\rangle\right)^{1/2}. 
\ee
The normalization is chosen such that $0\leq m_b \leq 1$\,.

\subsection{Bond Susceptibility}
Bond susceptibility is the response function associated with bond magnetization $M_b$
and its thermodynamic conjugate field $H_b$, called  bond ordering field.\cite{3}
The exact  physical nature of $H_b$ is not yet known to us; however, bond susceptibility can still
be calculated without appealing to a knowledge of $H_b$.
 
The change in  Gibbs free energy in an infinitesimal bond ordering process may be
written in terms of the newly introduced parameters $M_b$ and $H_b$, as in the following:
\be
dG = -SdT - M_b\,dH_b\;.\label{eq:one}
\ee
Eq.\,(\ref{eq:one})  can serve as  starting point for incorporating $M_b$ into the thermodynamics of 
disordered systems.
Bond susceptibility apart from a normalization will therefore be given by:
\be
\chi_b = \frac{\partial M_b}{\partial H_b} = -\frac{\partial^2 G}{\partial H_b^2}\;.
\label{eq:two}
\ee
Starting with Eq.\,(\ref{eq:two}), one may readily obtain  an expression 
for the bond susceptibility in terms of fluctuations in  the bond magnetization:
\be
\chi_b=\left(\langle |\M_b|^2 \rangle - \langle |\M_b|\rangle^2\right)/2Nk_BT\;, 
\label{eq:three}
\ee
where $\M_b=\sum_{<i,j>}^{2N} \bu_{ij}$\,.
Eq.\,(\ref{eq:three}) is properly normalized to the number of \nn\ bonds $2N$ 
for a system of $N$ atoms with periodic boundary conditions,
and is  used for  calculating  bond susceptibility  via MC simulations.

This rather abstract entity, the bond susceptibility, describes the tendency of a 
system for bond ordering and provides the basis for a new identification of  glass transition
temperature.

\subsection{New Identification for T$_{\mbox{g}}$ }
Following the previous discussion one can trace  the origins of  MRO
characteristic of the vitreous state in  the local ordering of bonds, which becomes most intense
at some particular temperature $T_g$.
This brings us to another identification for the  calorimetric glass transition temperature,
which particularly applies to fragile and intermediate class of the glass-forming liquids.\cite{4} 
In glass transition region,  bond susceptibility 
of a supercooled glass-forming liquid reaches a maximum.
This also implies that  specific heat must display a maximum in the glass transition region
because of  the large energy fluctuations associated with the intense ordering of bonds.
In view of this, the transition peak in the experimentally measured specific 
heat of various fragile and intermediate glass-formers can be regarded
as an artifact of intense bond ordering or strong fluctuations in the number 
of bonds in each energy state occurring at glass transition.
Hence, we propose a new identification for $T_g$ as that temperature corresponding to maximum 
of  bond susceptibility of a supercooled liquid nearing configurational arrest.

Having  given an  explanation for the specific heat  peak at the glass transition, 
it doesn't seem improper to  consider the unexpected linear behavior of specific
heat at very low temperatures.\cite{5}
Work by Hunklinger \etal\ gives strong evidence that the
low temperature anomalous properties of amorphous materials arise mainly from  two-level systems
and not from the multi-level vibrational degrees of freedom associated with the atoms.\cite{6} The energy gaps 
$\Delta$ of the two-level systems are supposed to vary with uniform probability in some range
$0\leq \Delta \leq \Delta_0$, where $\Delta_0\approx 1 K$.\cite{6.1,6.2} These  low energy excitations may be 
attributed to the  bond ordering process  at low temperatures.  In  disordered systems, 
whether supercooled liquids or glass, bonds continuously
relax into more stable internal energy states with lowering of the temperature.
At very low temperatures certain number of bonds may be seen to act like two-level systems with varying
energy gaps. Bond ordering process at low temperatures is therefore a possible explanation
for the low temperature anomalous properties  exhibited by  amorphous materials.

\subsection{Structural Magnetization and Susceptibility}
Structural magnetization is a   measure of the conventional LRO
a system may possess. For the system we are considering
the displacement of an atom from its designated equilibrium 
lattice site $i$, is characterized with  $\r_i$. Further idealizing the system by assuming that 
 displacements are of same magnitude $|\r |$, we can express  extensive 
structural magnetization  as in the  following:
\begin{eqnarray}
M_s & = & \left\langle \left( \sum_{i=1}^N \ru_i \right)^2 \right\rangle^{1/2}\;,
\end{eqnarray}
where $\hat{\r_i}=\r_i / |\r|$\,.
The  intensive (or normalized)  structural magnetization is  given by; $m_s=M_s/N$, 
where $0\leq m_s \leq 1$\,.
The analogy between  structural magnetization and  magnetization of a magnetic system, is
rather obvious. In fact, we can use this analogy to express  structural susceptibility 
in terms of fluctuations in  structural magnetization, as follows:
\be
\chi_s = \left(\langle |\M_s|^2\rangle - \langle |\M_s|\rangle^2\right)/Nk_BT \;, 
\ee
where $\M_s = (\sum_i R_i^x\;,\;\sum_i R_i^y)$\,. Structural susceptibility is the response function
describing the tendency for structural ordering.

\subsection{Order-Parameter for Glass}
We would like to address at this point a possible order parameter (or order parameter density)
for supercooled liquids and glass, that also serves as yet another distinction 
between fragile and strong classes of the glass-forming liquids. 
This will simply be the bond magnetization 
if one is  considering strictly a $\mbox{liquid}\!\leftrightarrow\!\mbox{glass}$ transition.

Many of the theories describing glass transition phenomenon, assume that there is a single parameter 
which characterizes glass. This assumption is believed to be 
inaccurate.\cite{7} Prigogine and Defay
have  shown that in general the ratio of the discontinuities in second-order thermodynamic quantities;
isothermal compressibility, heat capacity at constant pressure, and coefficient of thermal expansion:
\be
R = \frac{\Delta \kappa_T \Delta C_p}{T V (\Delta \alpha)^2}\;,
\ee
is equal to unity if a single order parameter characterizes the underlying thermodynamic transition,
but if more than one order parameter is involved, then $R>1$\,.\cite{8} The latter seems to describe most glasses.
In view of this, we consider a two parameter description of  glass  which involves two of the 
parameters described earlier, namely, the structural magnetization $m_s$ and the bond order parameter $m_b$.
We require  for amorphous solids that the structural LRO  vanishes while 
the bond magnetization  remains significantly large. The requirement of  the 
vanishing of  structural LRO, is meant to characterize the liquid-like attributes of the amorphous
systems. Yet large values for  bond order parameter is a solid-like attribute that should serve to
distinguish the glass  from the liquid phase. 

Clearly, in  case of strong glass-formers characterized with
strong covalent bonds, the values for  bond order parameter 
above and below the transition must be  quite comparable.
However, in case of  fragile systems the liquid undergoes
substantial bond ordering at the  transition mainly due  to the nondirectional
nature of their chemical bonds. As a result bond magnetization is expected 
to vary rather significantly for the fragile class. 
Originally, the labels strong and fragile 
were introduced to refer to the ability of a liquid in  withstanding
changes in  MRO with  temperature.\cite{4}
In context of bond-ordering model,  these labels will be referring to the ability of a
liquid to withstand changes with  temperature in  bond  magnetization\ $m_b$.
It is worthwhile to mention that in this scheme an ideal glass may be  characterized
as being maximally bond ordered which should also imply the  least possible energy.

\section{Simulation: \\ Procedure and Results}
Clock models have been investigated in some detail, 
both analytically and through numerical methods.\cite{9,10,11}
Previous Monte Carlo works, have examined the behavior of 2D clock systems with
various values of the parameter $q$. The important results are that for \mbox{$q\leq 4$}\ the system exhibits
one second-order transition, but for $q>4$\ two Kosterlitz-Thouless (KT) transitions\cite{12} are present.
The upper transition temperature is believed to have a value approximately equal 
to  the KT transition temperature for the continuous model,\cite{13}
$T_{c}\!=\!0.89\;J/k_B$. 
As $q$\ increases, the lower transition temperature ($\propto\!1/q^2$) approaches
zero, leaving just one KT transition for the 2D $XY$-model.

For  model Hamiltonian we consider Eq.\,(6), which  in fact is a bond-diluted version of  
$q$=$8$\ state clock model, involving \nn\ interaction with 
antibonding electronic state.
For our purpose, the  eight possible  orientations must be interpreted as the possible 
displacements of an atom from its designated equilibrium lattice site, 
with every atom's displacement being  the same if the system were in a ordered configuration.

The standard MC importance-sampling method was used to simulate the behavior of the system on
\mbox{$L\!\times\!L$} square lattices with periodic boundary conditions. We performed simulations
on lattices of size $L\!=\!12,\;20,\;32,\;\mbox{and}\;50$. Preliminary work was carried out 
on systems of size $L\!=\!12$.
In every case the temperature was lowered  in steps of $0.05\;J/k_B$, starting with  initial value\ $T\!=\!2.00\;J/k_B$. 
The system was initialized in  random configuration suitable for high temperatures where it
is known to be disordered. At every temperature system was allowed to equilibrate
through a few hundred Monte Carlo steps per site (MCS). The data points were
then acquired by averaging over 40,000 MCS. The data were accumulated in several bins, and
binned averages were  used to obtain  error estimates for the calculated  means 
and also to monitor the state of equilibrium. For most cases the estimated statistical
errors were less than $2\%$ of the calculated mean values.
In order to test our code, we simulated $q\!=\!6$ state
clock model and compared our results with the extensive 
literature available on the subject.\cite{9,10}
To our satisfaction the agreements were impressive. 
As a  note on the calculation of internal  energy,  the (possible) bonds were sorted out in
ascending order of  energy at every time step and were given bonding-electrons in that order.

Many of the quantities calculated have been already discussed and  given explicit 
mathematical definitions in Sec.\,IV\@. In addition,  specific heat was obtained 
from fluctuations in the total energy:
\be 
C = \left( \langle \sch^2 \rangle - \langle \sch \rangle^2 \right)/2Nk_BT^2\;. 
\ee
The bond-related quantities;  bond order parameter, bond susceptibility, internal
energy and specific heat are normalized to the number of \nn\ bonds $2N$, for
a system of size $N\!=\!L^2$ with periodic boundary conditions. On the other hand the structural properties, 
\ie\ structural magnetization and susceptibility are normalized to the system size $N$.
Results correspond to a hole concentration of 40\%.

Perhaps the most interesting feature of the results is 
the conspicuous  peak in  bond susceptibility, shown in Fig.\,2,  
which is  an indication of bond ordering,
uncorrelated with any  long-range structural ordering as is evident from  
structural magnetization and susceptibility seen in Figs.\,3, and 4.
Finite size effects are quite  evident. The temperature at which bond susceptibility
reaches its maximum value is 
estimated to be  $0.9\;J/k_B$, consistent with the KT transition temperature.
At exactly the temperature where bond susceptibility maximum occurs, one observes 
a maximum in  specific heat shown in Fig.\,5,
which should be attributed to large  energy fluctuations associated with the ordering of bonds. Indeed,
experimental measurements of heat capacity for covalently bonded fragile systems such as As$_2$S$_3$ and 
B$_2$O$_3$, exhibit similar  peaks at 
glass transition which therefore suggest  bond ordering nature for the glass transition. 
The internal energy for various system sizes is shown in Fig.\,6, displaying  steep slope
in bond ordering region. 

Fig.\,7 contains the variation of bond magnetization with temperature.
In bond ordering region,  bond magnetization increases rapidly with decreasing temperature
in spite of the fact that structural magnetization stays fairly constant there.
This behavior testifies to the earlier assertion that a system can undergo substantial 
bond ordering and hence largely reduce its internal energy without undergoing any 
significant structural ordering.

Unlike the behavior expected from $q$-state clock model in 2D, structural susceptibility 
(Fig.\,4) does not exhibit  singular behavior at intermediate temperature range hence
ruling out the possibility of  long-range cooperative structural ordering. 
This behavior of structural susceptibility may be seen in view of the large concentration 
of holes or bond dilution.

\section{Summary}
We have investigated the  existence of order associated with  bonds,
in  amorphous systems of interest such as vitreous silica.
Through our MC simulations, it is found that bond ordering may occur
independently of structural ordering.
Bond ordering  implies  ordering in  \nn\ and \nnn\ distances and thus
leads to MRO, which is a key aspect of the vitreous state.

An order parameter for  supercooled liquids and glasses is introduced. 
In  case of  $\mbox{liquid}\leftrightarrow\mbox{glass}$ transition bond magnetization $m_b$
may be fruitfully employed in the thermodynamics of glass and any calculation that may involve 
coarse-grained parameters. In addition, a new identification for glass transition 
temperature is afforded through the variation with temperature of the bond susceptibility.

As a last  remark, the sharp rise in the viscosity of glass-forming liquids
when cooled toward   glass transition,\cite{4}
may be viewed in terms of  strengthening of bonds.
The fragile glass-formers undergo significant  bond ordering through the transition 
and for that reason their viscosity rises dramatically,
as opposed to the smooth \mbox{Arrhenius-like} behavior of viscosity 
exhibited by  strong glass-forming liquids.

\begin{figure}
%\begin{center}
%\leavevmode
%\centerline{\epsfysize=6cm \epsfxsize=8.5cm \epsfbox{fig1.eps}}
%\end{center}
\caption{The distribution of Si-O-Si bond angle,  measured by Mozzi 
and Warren.\cite{1}}
\end{figure}

\begin{figure}
\begin{center}
\leavevmode
\centerline{\epsfysize=6cm \epsfxsize=8.5cm \epsfbox{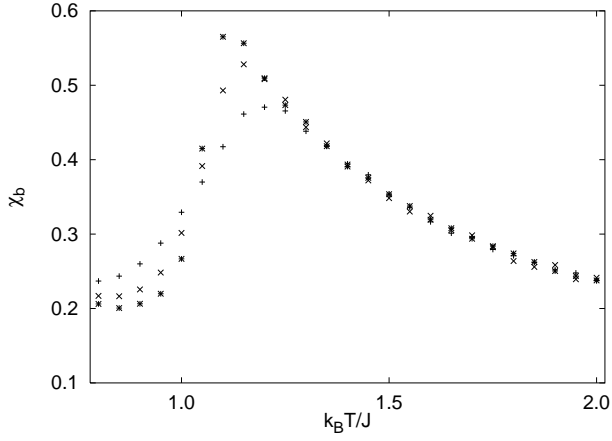}}
\end{center}
\caption{Plot of  bond susceptibility vs.\ temperature for three different system sizes;
$20^2$ (+), $32^2$ ($\times$), and $50^2$ ($\ast$).}
\end{figure}

\begin{figure}
\begin{center}
\leavevmode
\centerline{\epsfysize=6cm \epsfxsize=8.5cm \epsfbox{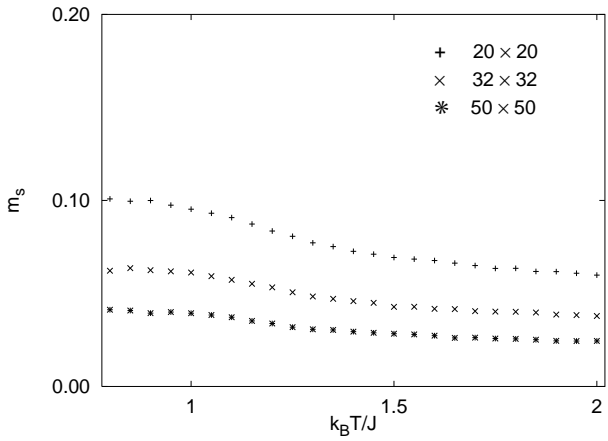}}
\end{center}
\caption{Plot of structural magnetization  vs.\ temperature. Each curve is 
normalized to the corresponding system size N.}
\end{figure}

\begin{figure}
\begin{center}
\leavevmode
\centerline{\epsfysize=6cm \epsfxsize=8.5cm \epsfbox{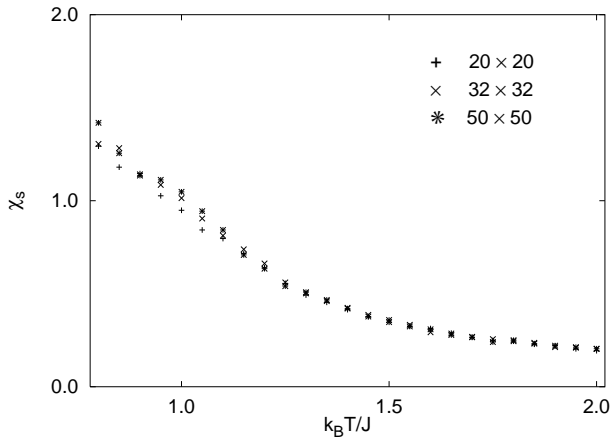}}
\end{center}
\caption{Variation of structural susceptibility with temperature. Clearly, there is
no indication of long-range structural ordering in the temperature range shown.}
\end{figure}

\begin{figure}
\begin{center}
\leavevmode
\centerline{\epsfysize=6cm \epsfxsize=8.5cm \epsfbox{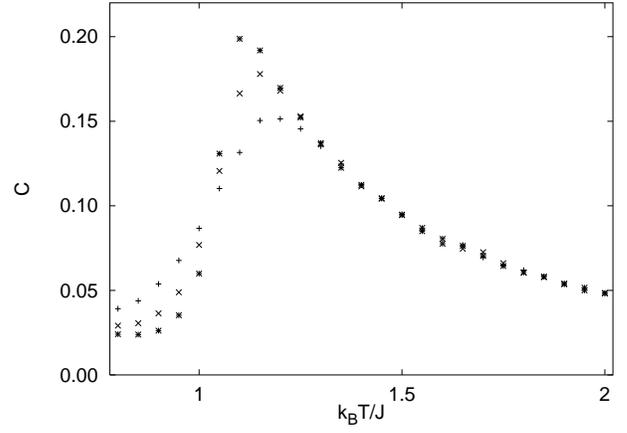}}
\end{center}
\caption{Specific heat as a function of temperature for three different system sizes;
$20^2$ (+), $32^2$ ($\times$), and $50^2$ ($\ast$).}
\end{figure}

\begin{figure}
\begin{center}
\leavevmode
\centerline{\epsfysize=6cm \epsfxsize=8.5cm \epsfbox{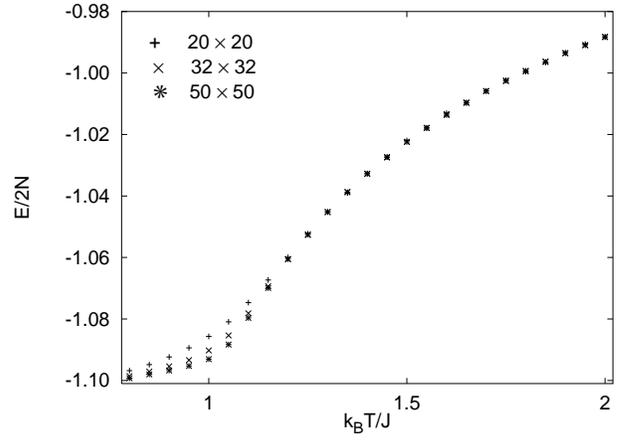}}
\end{center}
\caption{Energy per bond as a function of temperature. The total energy $E$ is in units
 of the coupling $J$.}
\end{figure}

\begin{figure}
\begin{center}
\leavevmode
\centerline{\epsfysize=6cm \epsfxsize=8.5cm \epsfbox{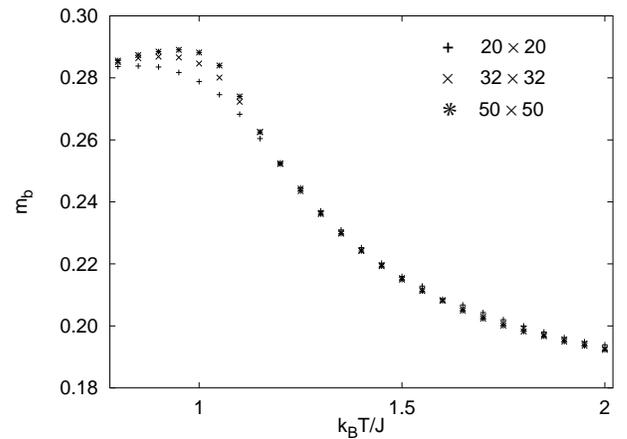}}
\end{center}
\caption{Plot of bond magnetization  vs.\ temperature. 
There  appears a sharp variation due to bond ordering, uncorrelated with any 
long-range structural ordering.}
\end{figure}


\begin{references}
\bibitem{3} A  field whose only effect is to lower the energies of bonds.
\end{references}
\end{document}